\documentclass[%
 preprint,
 amsmath,amssymb,
 aps,
prd,
]{revtex4-1}

\usepackage{graphicx,amsmath,float}% Include figure files
\usepackage{dcolumn}% Align table columns on decimal point
\usepackage{bm}% bold math
\usepackage{placeins}
\begin{document}

%\preprint{APS/123-QED}

\title{Two competing interpretations of Kelvin probe force microscopy on semiconductors put to test}

\author{Leo Polak}
  \email{l.polak@vu.nl}
% \altaffiliation[Also at ]{Physics Department, XYZ University.}
\author{Rinke J. Wijngaarden}
\affiliation{Division of Physics and Astronomy, VU University Amsterdam, Amsterdam, The Netherlands}

\date{\today}

%Abstract
\begin{abstract}
Kelvin probe force microscopy (KPFM) is a popular tool for studying properties of semiconductors. However, the interpretation of its results is complicated by the possibility of so-called band bending and the presence of surface charges. In this work we study two different interpretations for KPFM on semiconductors: the contact potential difference (CPD) interpretation, which interprets the measured potential as the work function difference between the sample and the probe, and a newer, alternative, interpretation proposed by Baumgart, Helm and Schmidt (BHS). By performing model calculations we demonstrate that these models generally lead to very different results.  Hence it is important to decide which one is correct. We demonstrate that BHS predictions for the Kelvin voltage difference between the \emph{p} and  \emph{n} parts of a \emph{pn}-junction are inconsistent with a set of experimental results from the literature. In addition, the BHS interpretation predicts an independence from the probe material as well as from surface treatments, which we both find to disagree with experiment. On the other hand, we present a theoretical argument for the validity of the CPD interpretation and we show that the CPD interpretation is able to accommodate all of these experimental results. Thus we posit that the BHS interpretation is generally not suitable for the analysis of KPFM on semiconductors and that the CPD interpretation should be used instead.
\end{abstract}

\maketitle

%Body of paper
\section{Introduction}
Kelvin probe force microscopy (KPFM) is an advanced atomic force microscope (AFM) method that enables the study of electrical properties of a sample with high lateral resolution. For semiconductor samples, these properties include the dopant density, density of surface states, surface charge density, band bending and the work function \cite{saraf2005localAPL,saraf2005localSS,tzeng2006charge,barbet2008surface,tsui2008two,volotsenko2010secondary,arita2014surface,maragliano2014quantifying}. In combination with sample illumination techniques, properties such as the band gap, carrier diffusion length, and recombination rate can be obtained \cite{kronik1999surface,meoded1999direct,streicher2009surface}. The importance of KPFM is reflected in its application in a broad range of popular material science topics, such as new photovoltaic materials \cite{watanabe2014situ,xiao2015giant,yun2015benefit,fuchs2016high}, two-dimensional materials \cite{lee2009interlayer,bussmann2011doping,kim2013work,dumcenco2015large}, nanowires \cite{koren2010measurement,jeong2014quantitative,gupta2015nanoscale}, 
topological insulators \cite{hao2013fermi}, plasmonic structures \cite{sheldon2014plasmoelectric,gwon2015plasmon}, and photocatalytic systems \cite{hiehata2007local,kittel2016charge}. Reviews of KPFM and its applications can, e.g., be found in Refs. \cite{melitz2011kelvin,sadewasser2012experimental}.

Like the classic vibrating Kelvin probe, the quantity measured with KPFM is generally interpreted as the contact potential difference (CPD) \cite{nonnenmacher1991kelvin,melitz2011kelvin,sadewasser2012experimental}. In the case of semiconductors, the situation is complicated by the possibility of band banding near the surface and the possible presence of surface charges.  Application of the CPD interpretation therefore requires careful consideration of these effects on the work function \cite{brattain1953surface,kronik1999surface,saraf2005localAPL,saraf2005localSS,tzeng2006charge,tsui2008two,barbet2008surface,volotsenko2010secondary}. However, Baumgart, Helm, and Schmidt \cite{baumgart2009quantitative,baumgart2011kelvin,baumgart2012quantitativethesis} proposed an alternative interpretation for KPFM on semiconductors, which we will refer to as the BHS interpretation. As we show below, the CPD and BHS interpretations are significantly different. Hence, it is important to determine what the differences between these two interpretations are, and to what extent they are valid \cite{kuriplach2014improved}. This is the main purpose of this work. 

This article is organized as follows. In the theory section, we introduce the principles of Kelvin probe measurements, describe the CPD and BHS predictions for \emph{pn}-junctions, and give the relevant expressions for the semiconductor modeling. In the methods sections we describe the computational and experimental details. Finally, in the results and discussion section we explore the general differences between the BHS and CPD interpretations and test them against experimental KPFM results.

\section{Theory}
\subsection{Kelvin probe principles} \label{KPprinciples}
In KPFM, an AFM probe and sample are electrically connected through a voltage source that applies an oscillating potential $V=V_\text{DC}+V_\text{AC}\text{ cos }\omega t$. This causes an oscillation of the electrostatic force per unit area at frequency $\omega$ with amplitude $F_{\omega}$, which is called the first harmonic. KPFM methods can be divided into two main categories: amplitude modulation (AM) and frequency modulation (FM). In closed loop AM-KPFM, $V_\text{DC}$ is adjusted by a feedback loop to the value $V_K$ that nullifies a signal that is proportional to $F_\omega$, i.e.,
\begin{equation}
\left. F_{\omega} \right\vert_{V_\text{DC}=V_{K}}=0.
\label{AMCondition}
\end{equation}
In closed loop FM-KPFM a signal is nullified that is approximately proportional to the amplitude of the first harmonic of the gradient of the electrostatic force \cite[][Eq. 2.18]{sadewasser2012experimental}, i.e.
\begin{equation}
\left. \frac{\partial F_{\omega}}{\partial z} \right\vert_{V_\text{DC}=V_{K}}=0.
\label{FMCondition}
\end{equation}
Because the subject of this work is the interpretation and modeling of the quantity $V_K$ obtained with KPFM on semiconductor samples we now introduce the theoretical background of the Kelvin voltage and of its interpretation in terms of the CPD and BHS models. 

Upon electric connection of two conducting bodies with different work functions, a potential difference $V_\text{CPD}$ is generated between their surfaces. The work function, $W$, of an object is defined as the energy to bring an electron from the bulk of the object to a position just outside its surface, in the absence of a net charge on the object and any external electric fields originating from other objects. $W$ can vary over the surface of an object with homogeneous bulk properties, because it contains contributions from potential drops at the surface, such as the band bending potential at the semiconductor surface.

To enable a simple theoretical discussion of KPFM measurements, we reduce the problem to one dimension. In this simplified configuration, the connected KPFM probe and sample are positioned opposite each other and form a parallel plate capacitor. Also, in this approximation, each has a single work function. Hence, 
\begin{equation}
V_\text{CPD}\equiv (W_s-W_p)/e,
\label{CPDWS}
\end{equation}
where $e$ is the positive elementary charge and $W_s$ and $W_p$ are the sample and probe work function, respectively.

First we consider the case of ideal conductors with surface properties that are independent of any applied potentials. In this case the charge on each body is proportional to the total potential difference. We define the feedback voltage positive when a positive voltage is applied to the sample with respect to the probe. The total net charge per unit area is then
\begin{equation}
\sigma_s=C\left(V-V_\text{CPD}\right).
\label{Qmetal}
\end{equation}
The proportionality constant $C$ is the capacitance per unit area. At a plate distance $z$, $C=\varepsilon/z$ and the electrostatic force per unit area is
\begin{equation}
F=\frac{\sigma_s^2}{2\varepsilon},
\label{Fideal}
\end{equation}
where $\varepsilon$ is the permittivity of the medium in the gap. The first harmonic is then equal to
\begin{equation}
F_{\omega}=\frac{\varepsilon}{z^2}\left(V_\text{DC}-V_\text{CPD}\right)V_\text{AC}.
\label{Fwideal}
\end{equation}
Clearly, in this ideal case we see from Eqs (\ref{AMCondition}) and (\ref{FMCondition}) that both AM- and FM-KPFM methods lead to
\begin{equation}
V_K=V_\text{CPD},
\label{CPDinterp}
\end{equation}
which is the CPD interpretation of $V_K$. In the three-dimensional KPFM geometry, this corresponds to interpreting the measured potential as an approximation for the difference between the work function in a small area of the sample directly underneath the tip and the work function of the tip apex of the probe.

In the case of a semiconducting sample (still probed with a metallic probe), the situation becomes more complicated. It is the main purpose of this paper to evaluate how KPFM results for this configuration should be interpreted. One main complication of a semiconducting sample is that electrical fields penetrate the sample and influence the charge distribution inside the sample. Equivalently, the conduction- and valence-bands of the semiconductor are generally at a different position close to the surface as compared with the bulk, which is the well-known band-bending at semiconducting surfaces \cite{kronik1999surface}. As discussed in more detail below, the surface band bending changes the total potential difference between the surfaces of the two bodies. Hence, $\sigma_s$ is not simply proportional to the sum of the applied potential and the work function difference, as in (\ref{Qmetal}). Instead, the charge-voltage relation can be described with a voltage dependent capacitance per unit area, $C(V)$, as
\begin{equation}
\sigma_s=\int_{V_\text{CPD}}^{V} C(V^{\prime })dV^{\prime }.
\label{Qsemi}
\end{equation}
As a result, Eq. (\ref{CPDinterp}) might not be valid for semiconducting samples and hence the CPD interpretation might be wrong. However, we now present a theoretical argument for its validity.

The electrostatic force for a voltage dependent capacitance can still be written as in Eq. (\ref{Fideal}) \cite{hudlet1995electrostatic}. Combining this expression with Eq. (\ref{Qsemi}) it is clear that without modulation of the potential, i.e. $V_\text{AC}=0$, $F$ will be zero when $V=V_\text{DC}=V_\text{CPD}$. However, this does not necessarily mean that with modulation $F_\omega$ will be nullified by $V_\text{DC}=V_\text{CPD}$. To solve this issue we use a similar approximation as was used by Hudlet et al. \cite{hudlet1995electrostatic}. We make a first order Taylor expansion of $F(V)$ around $V_\text{DC}$ and take the term proportional to $\text{cos }\omega t$ as an approximation for $F_\omega$. This leads to
\begin{equation}
F_{\omega}\approx V_\text{AC}\left.\frac{\partial F}{\partial V}\right\vert_{V=V_\text{DC}}.
\label{Fwfirstorder}
\end{equation}
With (\ref{Fideal}) and (\ref{Qsemi}) this becomes
\begin{align}
F_{\omega}&\approx \frac{V_\text{AC}}{2\varepsilon}\left.\frac{\partial}{\partial V}\left(\int_{V_\text{CPD}}^{V} C(V)dV\right)^2 \right\vert_{V=V_\text{DC}} \nonumber \\
&=\frac{V_\text{AC}}{\varepsilon}\left(I\left(V_\text{DC}\right)-I\left(V_\text{CPD}\right)\right)C\left(V_\text{DC}\right),
\label{Fwfirstorder2}
\end{align}
where $I(V)$ is the antiderivative of $C(V)$. Because $C(V)$ is always positive, $I$ is a monotonically increasing function. Hence, this approximation for $F_{\omega}$ is only nullified by $V_\text{DC}=V_\text{CPD}$. This indicates that the CPD interpretation given by Eq. (\ref{CPDinterp}) is (despite a voltage dependence of $C$) valid for KPFM on semiconductors. For FM-KPFM, there is just a $\partial / \partial z$ added in front of Eq. (\ref{Fwfirstorder}), see Eq. (\ref{FMCondition}). Therefore, with the same reasoning, Eq. (\ref{CPDinterp}) would also be valid for FM-KPFM.

In principle, higher order contributions to $F_{\omega}$ can shift $V_{K}$ from $V_\text{CPD}$, but this can be avoided by keeping $V_\text{AC}$ small. A more precise analysis of this effect requires careful consideration of the frequency dependent dynamics of the surface state charge and the space charge layer, which is outside the scope of the present work.

Baumgart et al. \cite{baumgart2009quantitative,baumgart2011kelvin,baumgart2012quantitativethesis} argued that the CPD interpretation is invalid for semiconductors and proposed the alternative BHS interpretation. To further investigate the merits of both interpretations, we evaluate them for a well defined situation: the potential difference $\Delta V_{K}$ between the \emph{p}- and \emph{n}-sides of \emph{pn}-homojunctions as measured by KPFM:
\begin{equation}
\Delta V_{K} \equiv V_{K,p}-V_{K,n},
\label{DeltaVK}
\end{equation}
where the subscripts $p$ and $n$ indicate that the values are evaluated on the \emph{p}- and \emph{n}-type areas, respectively. In the next two subsections we calculate $\Delta V_{K}$ for the CPD and BHS interpretations.

\subsection{\emph{pn}-junctions in the CPD interpretation}
According to the CPD interpretation:
\begin{equation}
e\Delta V_{K}^\text{CPD}=W_{s,p} - W_{s,n},
\label{pnCPD}
\end{equation}
which is clearly independent of the probe work function. Therefore, we only need to consider the semiconductor work function in the modeling.

\begin{figure}[t]
\includegraphics[width = 220pt]{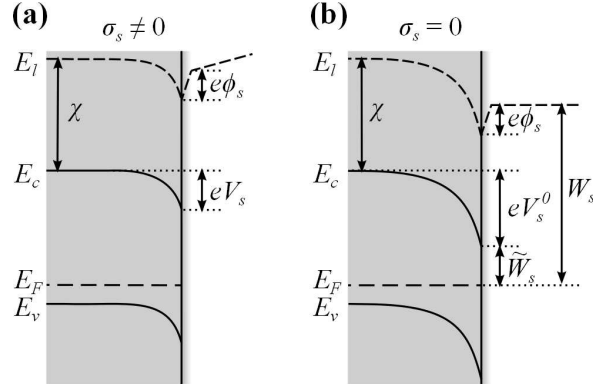}
\caption{\label{BandDiagram}Energy diagram of a semiconductor with (a) a nonzero net charge and (b) a zero net charge. Note that the work function $W_s$ and the related quantity $\widetilde{W}_s$ are defined in the uncharged condition.}
\end{figure}

Fig. \ref{BandDiagram} shows schematic energy level diagrams of a \emph{p}-type semiconductor with bulk at the l.h.s. and surface at the r.h.s. in the diagrams. Diagram (a) corresponds to a nonzero net charge and (b) to a zero net charge on the semiconductor. As is usual in such energy diagrams, the electron energy increases towards the top of the figure, hence electric potential increases towards the bottom. $E_{F}$ is the Fermi level in the semiconductor and $E_{v}$ and $E_{c}$ are the valence and conduction band energies in the bulk of the semiconductor, respectively. In the presence of surface charges or an external electric field, a so-called space charge region with non-zero net charge forms in the semiconductor just below the surface. This results in a potential difference, $V_{s}$, between the bulk and the surface of the semiconductor, which is called the band bending potential. In addition to the band bending, there is usually a potential step $\phi_{s}$ at the surface of the semiconductor due to a fixed dipole layer on the surface, which can be caused by surface termination or a molecular layer adhered to the surface. We will assume that $\phi_{s}$ is independent of any external electric fields and also that it is equal for the \emph{p}- and \emph{n}-side of the \emph{pn}-junction. $E_l$ is the local vacuum level, defined (following Marshak \cite{marshak1989modeling}) as the energy of an electron at a given point if it were at rest and free from the microscopic potentials of the crystal atomic lattice, but not free from the macroscopic potentials, such as those generated at surfaces or interfaces. The bulk electron affinity, $\chi$, is defined here as the energy required to bring an electron from the conduction band $E_c$ to the local vacuum level in the bulk of the material.
 
In Fig. \ref{BandDiagram}(b) the semiconductor has zero net charge, but there is still a band bending. This means that there is charge in the space charge region, which is compensated by surface charges. We label the band bending potential in this uncharged situation with $V_s^0$. For this zero net charge case, the work function $W_s$ is the energy to bring an electron from the Fermi level, $E_F$, to the local vacuum level $E_l$ outside the semiconductor. This leads to $W_s=E_c-E_F+\chi-e\phi_s- eV_s^0$ (note that in the figure $V_s^0$ is positive, while $\phi_s$ is negative). We define
\begin{equation}
\widetilde{W}_s=E_c-E_F-eV_s^0,
\label{EFSurf}
\end{equation}
which is the energy difference between the Fermi level and the conduction band at the surface. Since it is assumed that the fixed surface dipole layer, $\phi_s$ is equal on both sides of the \emph{pn}-junction, the CPD interpretation for $\Delta V_{K}$ on the \emph{pn}-junctions (\ref{pnCPD}) becomes
\begin{align}
e\Delta V_{K}^\text{CPD}=&\widetilde{W}_{s,p}-\widetilde{W}_{s,n}.
\label{pnCPD2}
\end{align}
Hence, $\Delta V_{K}$ can be obtained from the positions of the conduction band level in the bulk with respect to the Fermi level, and $V_s^0$.

\subsection{\emph{pn}-junctions in the BHS interpretation}
We quote the main part of the argument for the BHS model for KPFM on semiconductors from \cite{baumgart2009quantitative} ``\emph{In order to minimize the electrostatic force $F_{el}$ onto the probe, the asymmetric electric-dipole layer has to be removed. This is achieved by injecting majority charge carriers into the surface region in order to screen the unscreened immobile ionized dopant atoms. The charge neutrality condition is only fulfilled when surface states discharge simultaneously}.'' Supposedly, on \emph{n}-type semiconductors this is achieved by applying a potential equal to \cite[][p. 40-41]{baumgart2012quantitativethesis}
\begin{equation}
eV_{K}^\text{BHS}=E_{c}-E_{F} \qquad \text{(\emph{n}-type)}
\label{Baumn}
\end{equation}
and on \emph{p}-type
\begin{equation}
eV_{K}^\text{BHS} =E_{v}-E_{F} \qquad \text{(\emph{p}-type)}.
\label{Baump}
\end{equation}
In addition, they expect a sample specific potential offset that is, according to Baumgart et al. \cite{baumgart2009quantitative}, independent of the work function of the probe. As a result, this interpretation predicts that $V_K$ is independent of the probe work function, which directly contradicts the CPD interpretation (Eq. (\ref{CPDinterp}) with Eq. (\ref{CPDWS})), which depends linearly on the probe work function.

Direct application of Eqs. (\ref{Baumn}) and (\ref{Baump}) would result in negative values for $\Delta V_{K}$, while in Refs. \cite{baumgart2009quantitative,baumgart2011kelvin,baumgart2012quantitativethesis} they state only positive values. This is achieved by taking absolute values as described in Ref. \cite[][p. 44]{baumgart2012quantitativethesis}. The resulting expression can be written as
\begin{equation}
e\Delta V_{K}^\text{BHS}=E_{c,n}-E_{v,p}.
\label{pnBaum}
\end{equation}

We note that, a priori, there appear to be some issues with the BHS model. In the one-dimensional description, even an asymmetric dipole layer at the semiconductor surface does not cause an electrostatic force. Hence, the argument, that the dipole layer has to be removed in order to minimize the electrostatic force, seems to be invalid, unless taking into account the real geometry somehow justifies this assumption. At the same time, they state as a second condition that charge neutrality has to be fulfilled, which is also the condition underlying the CPD interpretation. However, it is unclear how these two conditions are met simultaneously by Eqs. (\ref{Baumn}) and (\ref{Baump}). In addition, they apparently neglect the 'bulk work function difference', i.e. the work function difference minus the surface contributions, but do not mention why this is allowed. On the other hand, the BHS model seems to work well for the experiments analyzed by Baumgart et al. \cite{baumgart2009quantitative,baumgart2011kelvin,baumgart2012quantitativethesis}, hence it is important to further discuss and test its validity, which we do below.

\subsection{Semiconductor modeling} \label{models}
To predict $\Delta V_K$, we need to find the position of the band edges in the bulk with respect to the Fermi level (for CPD and BHS) and the zero net charge band bending potential $V_s^0$ (for CPD only).

For a non-degenerate \emph{n}-type semiconductor, the \emph{ position of the band edges in the bulk with respect to the Fermi level} can be approximated by solving \cite{sze2006physics} 
\begin{equation}
N_{c} \text{exp} \left( - \dfrac{E_{c}-E_{F}}{kT} \right) \approx \dfrac{N_{D}}{1+g_{D}\text{exp}[(E_{F}-E_{D})/kT]},
\label{EFermi}
\end{equation}
where $N_{c}$ is the effective density of states in the conduction band, $N_{D}$ is the donor concentration, $E_{D}$ is the donor level energy and $g_{D}$ is the ground state degeneracy of the donor level. A similar expression can be used for a \emph{p}-type semiconductor.

The \emph{zero net charge band bending potential}, $V_s^0$, is the value of $V_s$ for which the total net charge on the semiconductor is zero, i.e. $\sigma_{s}=0$. $\sigma_{s}$ is the sum of the net charge in the space charge layer $\sigma_{sc}$ and the surface charges. Two types of surface charge densities can be distinguished: a surface state charge density, $\sigma_{ss}$, which depends on the energy between the Fermi level and the band edges at the surface, and a fixed surface charge density, $\sigma_{sf}$. Thus
\begin{equation}
\sigma_{s}=\sigma_{sc}+\sigma_{ss}+\sigma_{sf}.
\label{Sigs}
\end{equation}
For simplicity, we will only use models with either surface states or fixed surface charge, not both at the same time. To be able to do calculations, expressions for the three contributions to $\sigma_{s}$ are needed. In the remainder of this sub-section we discuss these contributions.

We start with the dependence of the \emph{space charge density} $\sigma_{sc}$ on the band bending potential $V_s$. The relation between $V_{s}$ and $\sigma_{sc}$ for a \emph{p}-type semiconductor can be approximated by \cite{sze2006physics}
\begin{equation}
\sigma_{sc}=-\text{sgn}\left[V_{s}\right] \sqrt{2 \varepsilon_s N_{A} k T}G(V_{s}),
\label{Sigsc}
\end{equation}
where $\varepsilon_s$ is the permittivity of the semiconductor, $N_{A}$ is the acceptor concentration, $k$ is the Boltzmann constant, $T$ is the temperature, and
\begin{equation}
G=\sqrt{\text{exp}[-\beta V_{s}]+\beta V_{s}-1+\frac{n_e}{n_h}(\text{exp}[\beta V_{s}]-\beta V_{s}-1)},
\label{G}
\end{equation}
where $\beta=e/kT$, and, respectively, $n_e$ and $n_h$ are the equilibrium electron and hole carrier densities in the bulk. In addition, for non-degenerate \emph{p}-type semiconductors one can use the approximation $n_e/n_h\approx n_{i}^{2}/N_{A}^{2}$, where $n_{i}$ is the intrinsic carrier density. Similar expressions can be used for a \emph{n}-type semiconductor.

Now we discuss the dependence of the variable \emph{surface state charge density} $\sigma_{ss}$ on the band bending potential $V_s$. Surface states can be donor or acceptor type. Just as the states in the conduction and valence band near the surface, they are shifted by band bending. According to Fermi-Dirac statistics, the charge in acceptor surface states can be written as
\begin{equation}
\sigma_{ss}^A=\int_{E_{v}}^{E_{c}} \dfrac{-en_{ss}^A(E)}{1+\text{exp}[(E-E_{F}-eV_{s})/kT]}dE.
\label{Sigss}
\end{equation}
where $n_{ss}^A(E)$ is the acceptor density of surface states (DOSS) (per unit area and energy) in case of zero band bending, ignoring surface state degeneracy. A similar expression can be used for donor surface states. We will label the combination of donor and acceptor DOSS with $n_{ss}(E)$ and the total number of surface states with $N_{ss}$. 

On atomically clean Si, the total number of states $N_{ss}$ can be on the order of the density of surface atoms \cite{allen1962work}, i.e. $10^{15}$~cm$^{-2}$, while on hydrogen terminated Si surfaces it can be as low as $10^{10}$~cm$^{-2}$ \cite{flietner1988spectrum}. Significant variations in the functional dependence $n_{ss}(E)$ on Si have been reported \cite{hasegawa1983electrical}. Often, it is considered to have a U-shape, with acceptor states above and donor states below the minimum density \cite{hasegawa1983electrical,flietner1988spectrum}. However, Gaussian \cite{ragnarsson2000electrical,saraf2005localAPL}, Lorentzian \cite{volotsenko2010secondary}, delta \cite{kronik1999surface,monch2001semiconductor} and constant \cite{bardeen1947surface} functions have also been considered.

\begin{figure}[t]
\includegraphics[width = 220pt]{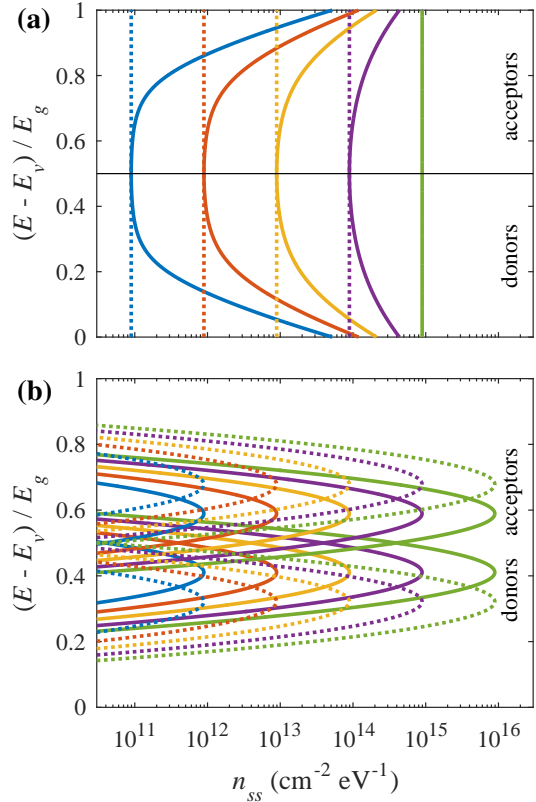}
\caption{\label{SurfaceStateDist}Model densities of surface states, $n_{ss}(E)$, used in Eq. (\ref{Sigss}). (a) U-shaped (solid lines) and constant (dotted lines) densities consisting of donor states in the lower half of the band gap and of acceptor states in the upper half. (b) Double Gaussian densities, which have 0.04~eV standard deviation and are centered at $E_{g}/2\pm0.1$~eV (solid lines) and $E_{g}/2\pm0.2$~eV (dotted lines). The Gaussians centered below $E_{g}/2$ consists of donor states and the Gaussians centered above $E_{g}/2$ consists of acceptor states. The constant and Gaussian densities in blue, red, orange, purple and green (from the left to right) correspond, respectively, to $N_{ss}=10^{11}$, $10^{12}$, $10^{13}$, $10^{14}$ and $10^{15}$~cm$^{-2}$. The U-shaped densities have the same $n_{ss}$ at $E_g/2$ as the constant densities with the same color, but higher $N_{ss}$.}
\end{figure}

To capture the main phenomenology of $n_{ss}\left( E\right)$, we consider three types of DOSS: U-shaped, constant and double Gaussian densities. Fig. \ref{SurfaceStateDist}(a) shows examples of the U-shaped (solid lines) and constant (dotted lines) densities. These consist of donor states in the lower half of the band gap and of acceptor states in the upper half. The U-shaped densities were chosen similar to those presented in Ref. \cite{flietner1988spectrum} for Si/SiO$_{2}$ interfaces with various surface treatments (in particular, for these curves we used $n_{ss}(E)=\alpha \text{ exp}[(E-\beta)^2/\gamma]+\delta$). Fig. \ref{SurfaceStateDist}(b) shows examples of the double Gaussian densities, which have 0.04 eV standard deviation and are centered at $E_{g}/2\pm0.1$~eV (solid lines) and $E_{g}/2\pm0.2$ eV (dotted lines). The Gaussian densities centered below $E_{g}/2$ represent donor states, while those centered above $E_{g}/2$ represent acceptor states. In the region close to the center of the band gap the Gaussian densities are similar to the results obtained by Angermann \cite{angermann2005interface} on an HF etched Si surface. Due to the symmetry of these DOSS models, $\sigma_{ss}$ will be zero when, at the surface, the Fermi level is in the center of the gap.

Finally, we will discuss the \emph{fixed surface charge density $\sigma_{sf}$}. Fixed surface charge is known to intrinsically exist at the Si/SiO$_{2}$ interface and to depend on specific sample treatments \cite{deal1967characteristics}. In addition, ions from the environment can deposit on the surface during sample preparation or during measurements \cite{brattain1953surface,okorn1999characterization}. In an experiment the value of $\sigma_{sf}$ is therefore often unknown. Negative fixed surface charge densities are not often considered, but for completeness we will also consider this possibility. Deposited ions can penetrate the native SiO present on the Si surface or remain on top of it. In our analysis below, however, we will neglect a possible distance between the Si surface and fixed surface charges.

\section{Methods}\label{methods}
\subsection{Computational methods}\label{CompMethods}
Our calculations according to the BHS interpretation, given by Eqs. (\ref{Baumn}) to (\ref{pnBaum}), only require knowledge of the position of the band edges in the bulk with respect to the Fermi level. This is calculated by numerically solving Eq. (\ref{EFermi}). Calculations according to the CPD interpretation additionally require computation of $V_s^0$. This is done by taking a model DOSS, $n_{ss}(E)$, or a fixed surface charge, $\sigma_{sf}$, and numerically solving $\sigma_s=0$ for $V_s$, using Eqs. (\ref{Sigs}) to (\ref{Sigss}).

We consider five surface models for fitting the CPD interpretation to experimental $\Delta V_K$ obtained on Si \emph{pn}-junctions: a constant $n_{ss}(E)$ with acceptor states in the upper half of the band gap and donor states in the lower half (labeled hereafter as `constant'), a $n_{ss}(E)$ with Gaussian distributed acceptor and donor states centered, respectively, at $\mu=E_{g}/2\pm0.1$~eV and standard deviation of 0.04~eV (labeled `Gauss1'), a $n_{ss}(E)$ with Gaussian distributed acceptor and donor states centered, respectively, at $\mu=E_{g}/2\pm0.2$ and also standard deviation of 0.04~eV (labeled `Gauss2'), a positive $\sigma_{sf}$ (labeled `$\sigma_{sf}>0$') and a negative $\sigma_{sf}$ (labeled `$\sigma_{sf}<0$'). We assume that the DOSS or fixed surface charge is the same on both sides of the \emph{pn}-junctions. For a given set of Si bulk parameters, the remaining fit parameter for the Gaussian and constant DOSS models is then $N_{ss}$ and for the fixed surface charge models it is $\sigma_{sf}$. Fitting was performed through iterative adjustment of the fit parameter, until the calculated $\Delta V_K$ was within 1 mV of the experimental value. We have not used the U-shaped DOSSs for fitting to experimental results, because, as shown below, its results are very similar to a constant DOSS, which is more simple to use.

For the fixed Si parameters, we use the values from Ref. \cite{sze2006physics}. These are: $\varepsilon_s=1.05\times 10^{-10}$~F/m, $E_g=1.12$~eV, $N_v=2.65\times 10^{19}$~cm$^{-3}$, $N_c=2.8\times 10^{19}$~cm$^{-3}$, $n_i=9.65\times 10^{9}$~cm$^{-3}$, $g_D=2$, $g_A=4$, $E_D(\text{P})=E_v+1.075$~eV, $E_D(\text{As})=E_v+1.066$~eV, and $E_A(\text{B})=E_v+0.045$~eV. In addition, we assume $T=293$~K.

\subsection{Experimental methods}
According to the BHS model, given by Eq. (\ref{Baumn}) and (\ref{Baump}), $V_K$ does not depend on the probe work function. To test this, we performed KPFM measurements in air with a Multimode 8 SPM with Nanoscope V controller and Signal Access Module (SAM) (Bruker) with four different probes on \emph{p}-Si, \emph{n}-Si, and Au. For each scan line the topography was first determined with standard tapping mode using amplitude feedback and then retraced with an offset (lift height) of 100~nm while performing closed loop AM-KPFM with excitation at the resonance frequency. Crosstalk was removed by external wiring of the excitation signal \cite{polak2014note}. The measurements were performed in dark, except for the laser beam used for detecting the probe deflection (1 mW, maximum at 690 nm). This beam illuminates the probe from the back side, such that the sample area close to the tip of the probe is shaded from direct illumination.

Si samples were cut from a single side polished \emph{p}-type $<$100$>$ wafer with $~5\times 10^{15}$~cm$^{-3}$ B dopant concentration and a single side polished \emph{n}-type $<$100$>$ wafer with $~1\times 10^{15}$~cm$^{-3}$ P dopant concentration. Before cutting, proper electric contact was created on the unpolished side of the wafers. This was done by first removing the native oxide layer through immersion in 1\% aqueous HF, followed by a quick rinse with demineralized (DI) water and drying under nitrogen flow, and then depositing 500 nm Al. On the \emph{n}-type wafer the contact side was additionally \emph{n}$^+$ doped prior to Al deposition. After making the contacts, the wafers were immersed in 1\% aqueous HF for 10 s, quickly rinsed with DI water, dried under nitrogen flow, and then stored in air.

The Au sample was created by magnetron sputtering 100~nm of Au on glass. As probes we used a gold coated probe (HQ:NSC14/Cr-Au, Micromasch), a PtIr coated probe (SCMPIT, Bruker), a TiN coated probe (FMG01/TiN, NT-MDT) and a special KPFM probe, which consists of a silicon tip on a silicon nitride cantilever with proprietary reflective (and conductive) back side coating(PFQNE-AL, Bruker).

After every two scans the probe was changed and the next probe was put in the same location with roughly 50~$\mu$m accuracy, using an optical microscope with top view. When each probe had been installed and used for taking two scans twice, the next sample was installed and the procedure was repeated. On the \emph{p}-Si sample measurements were performed on two different spots.

\section{Results and discussion}
\subsection{Comparing the BHS and CPD interpretation}
In this section we study predictions of the BHS and CPD interpretation according to the semiconductor models described in section \ref{models} for a wide range of dopant concentrations.

\begin{figure}[t]
\includegraphics[width = 220pt]{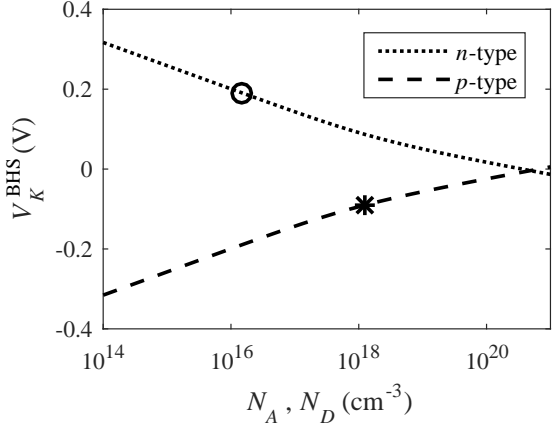}
\caption{\label{BaumFig}Variation of $V_K$ as a function of dopant concentration according to the BHS interpretation. The circle and star are described in the text as an example \emph{pn}-junction.} 
\end{figure}

Fig. \ref{BaumFig} shows the variation of $V_K$ on Si as a function of dopant concentration according to the BHS interpretation \footnote{Fig. \ref{BaumFig} is similar to the schematic diagram shown by Baumgart \cite[][Fig. 5.7]{baumgart2012quantitativethesis}.}. The dotted line corresponds to \emph{n}-type P-doped Si and the dashed line to \emph{p}-type B-doped Si. The expected value of $\Delta V_K$ for any Si \emph{pn}-junction can be read from Fig. \ref{BaumFig}. For example, consider a Si \emph{pn}-junction with $N_A$(B)$=1.27\times 10^{18}$~cm$^{-3}$ and $N_D$(P)$=1.49\times 10^{16}$~cm$^{-3}$. The values of $V_{K}$ on the \emph{p}-type and \emph{n}-type side are $-90$~mV and 191~mV and are indicated in Fig. \ref{BaumFig} with a star and circle, respectively. Hence, the predicted $\left\vert\Delta V_K\right\vert$ is 281~mV. Interestingly, from Fig. \ref{BaumFig}, the BHS interpretation predicts a general trend of decreasing $\Delta V_K$ with increasing dopant concentrations.

\begin{figure}[t]
\includegraphics[width = 220pt]{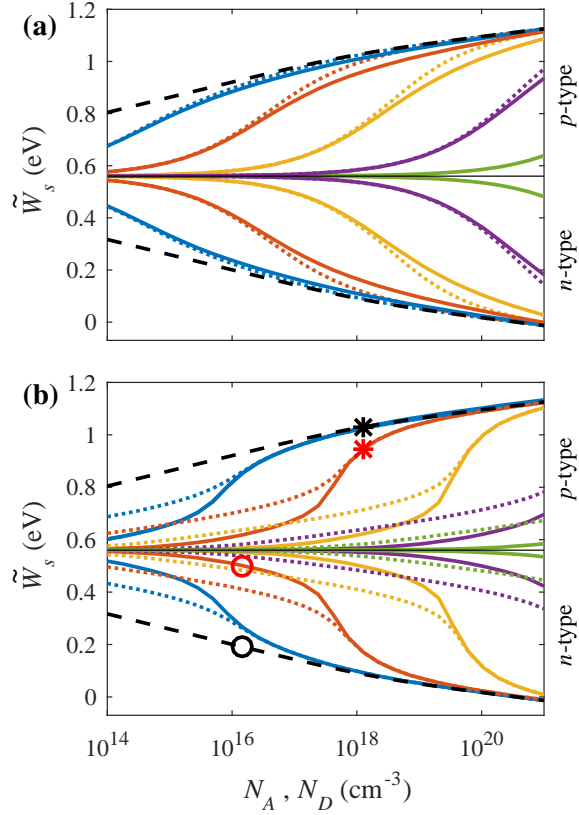}
\caption{\label{SurfaceFermi}$\widetilde{W}_{s}$ as a function of dopant concentration, calculated for Si with $\sigma_{sf}\equiv0$ and several different surface state distributions, $n_{ss}(E)$. The lines in the upper half of each figure correspond to B-doped \emph{p}-type Si and the lines in the lower half to P-doped \emph{n}-type Si. The black dashed lines correspond to zero $n_{ss}$ and, hence, $V_s=0$. The other results in (a) and (b) correspond respectively to the $n_{ss}$ shown in Fig. \ref{SurfaceStateDist}(a) and (b) with the same color and linestyle. The expected value of $\Delta V_K$ in the CPD interpretation for any Si \emph{pn}-junction with these $n_{ss}$ can be obtained from this data using Eq. (\ref{pnCPD2}). The black and red circles and stars in (b) are described in the text as an example \emph{pn}-junction.}
\end{figure}

In the CPD interpretation $\Delta V_K$ can conveniently be expressed in terms of $\widetilde{W}_{s}$, see eq. (\ref{pnCPD2}), hence we present the results of our calculations in terms of this quantity. Fig. \ref{SurfaceFermi}(a) and (b) show $\widetilde{W}_{s}$ as a function of dopant concentration, calculated for Si with $\sigma_{sf}=0$ and the various model DOSSs shown in Fig. \ref{SurfaceStateDist}(a) and (b), respectively, using identical line colors and types. The lower half of each sub-figure corresponds to \emph{n}-type P-doped Si and the upper half to \emph{p}-type B-doped Si. The black dashed lines correspond to zero band bending, i.e. $V_s=0$, which is the case when there are no surface states.
 
For each  DOSS shown in Fig. \ref{SurfaceStateDist}, the expected value of $\Delta V_K$ for any Si \emph{pn}-junction can read from Fig. \ref{SurfaceFermi} using Eq. (\ref{pnCPD2}). For example, consider again the same Si \emph{pn}-junction as above. Assuming a DOSS equal to the solid red line in Fig. \ref{SurfaceStateDist}(b), which has $N_{ss}=10^{12}$~cm$^{-2}$, the values of $\widetilde{W}_{s}$ on the \emph{p}-type and \emph{n}-type side are 946~meV and 501~meV and indicated in Fig. \ref{SurfaceFermi}(b) with a red star and circle, respectively. Hence, the predicted $\Delta V_K$ is 445~mV. In case of absence of surface states and fixed surface charge there would be zero band bending and $\widetilde{W}_{s}$ of the \emph{p}-type and \emph{n}-type side would lie on the black dashed lines as indicated by the black star and circle, respectively. In this case, the predicted $\Delta V_K$ would be 839~mV.

In a naive approach to the CPD interpretation, the band bending could be ignored, which corresponds to using the black dashed lines in Fig. \ref{SurfaceFermi}. Our calculations show for which range of parameters band bending is significant and, hence, where this naive approach fails. It clearly fails where $\widetilde{W}_{s}$ is close to the $ E_g/2$ and approximately independent of the doping concentration. This regime corresponds to the type of Fermi level pinning that was first suggested by Bardeen \citep{bardeen1947surface}, where $V_s^0$ can be approximated by the value of $V_s$ at which $\sigma_{ss}=0$ (instead of $\sigma_s=0$). For our symmetric model DOSSs this leads to $\widetilde{W}_{s}=E_g/2$.

\begin{figure}[t]
\includegraphics[width = 220pt]{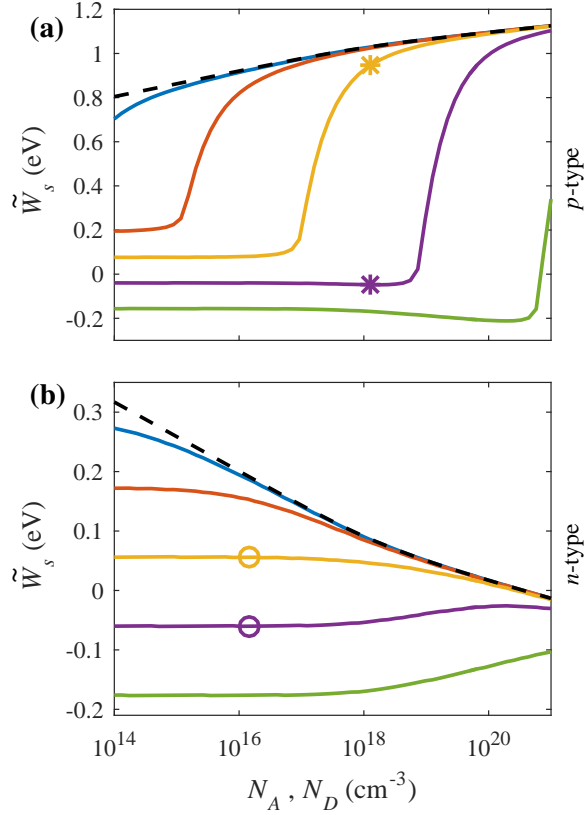}
\caption{\label{SurfaceFermi_Qsf}$\widetilde{W}_{s}$ as a function of dopant concentration, calculated for Si with $n_{ss}\equiv 0$ and $\sigma_{sf}/e=10^{10}$, $10^{11}$, $10^{12}$,$10^{13}$ and $10^{14}$~cm$^{-2}$ in blue, red, orange, purple and green (from top to bottom), respectively. The lines in (a) correspond to B-doped \emph{p}-type Si and the lines in (b) to P-doped \emph{n}-type Si. The black dashed lines correspond to zero $\sigma_{sf}$ and, hence, $V_s=0$.} 
\end{figure}

Fig. \ref{SurfaceFermi_Qsf} shows $\widetilde{W}_{s}$ as a function of dopant concentration, calculated for Si with positive fixed surface charge densities between $\sigma_{sf}/e =10^{10}$~cm$^{-2}$ and $10^{14}$~cm$^{-2}$. Subfigure (a) corresponds to \emph{p}-type B-doped Si and (b) to \emph{n}-type P-doped Si. The black dashed lines correspond again to zero band bending, which is the case when there is no fixed surface charge. Clearly, $V_s$ is positive for these surface charges. Negative fixed surface charge densities lead to similar results, but with opposite sign of $V_s$ and \emph{p}- and \emph{n}-type reversed.
 
From Fig. \ref{SurfaceFermi_Qsf} it is clear that a fixed surface charge density can have a dramatic influence on the work function and, therefore, also on $\Delta V_K$. To illustrate this, we consider again the same \emph{pn}-junction as above. For $\sigma_{sf}/e=10^{12}$~cm$^{-2}$, indicated with an orange circle and star, we obtain $\Delta V_K=944$~mV, while for $\sigma_{sf}/e=10^{13}$~cm$^{-2}$, indicated with a purple circle and star, we obtain $\Delta V_K=12$~mV, which is dramatically smaller. However, it should be noted that these calculations are less accurate for $\widetilde{W}_{s}<0$ and $\widetilde{W}_{s}>E_g$, because then the Boltzmann statistics assumed in Eqs. (\ref{EFermi}) and (\ref{Sigsc}) is less accurate. This is the case in the example with $\sigma_{sf}/e=10^{13}$~cm$^{-2}$, where $\widetilde{W}_{s}<0$ on both the \emph{p}- and the \emph{n}-side. Nevertheless, on both sides the Fermi level can be expected to be slightly above the conduction band edge at the surface, i.e. $\widetilde{W}_{s}$ is slightly below zero, and thus $\Delta V_K$ can be expected to be very small. Hence, the conclusion that $\Delta V_K$ is much smaller for $\sigma_{sf}/e=10^{13}$~cm$^{-2}$ than for $\sigma_{sf}/e=10^{12}$~cm$^{-2}$ still holds.

From these calculations it is clear that the CPD interpretation with our model DOSSs or a fixed surface charge density generally gives results that are significantly different from the BHS interpretation. The trend of decreasing $\Delta V_K$ for increasing dopant concentration found for the BHS interpretation is reversed in the case of the CPD interpretation with a DOSS. (In the case of a fixed surface charge density, the situation is more complicated.) As a result, it is not possible that both interpretations are correct and therefore it is important to settle this issue. In the next section we compare both interpretations with experiment.

\subsection{Testing the BHS interpretation against experiment}
We stress that the BHS interpretation does not depend on surface properties. As a result, when the dopant concentrations of a Si \emph{pn}-junction sample are given, there are no free parameters and the model directly predicts $\Delta V_K$. Although this is a very powerful feature, the observation of any deviation between predictions and experimental results would directly indicate that the interpretation is not correct. Therefore, to test the BHS interpretation, we now compare its predictions to experiments.

Table \ref{tab:table1} lists ten experimental values of $\Delta V_K$ obtained on Si \emph{pn}-junctions. The first column gives the corresponding references and second column labels each case for future reference. The third and fourth column state the dopant concentrations and dopant types of the two sides of the \emph{pn}-junction. The fifth column lists the experimental values of $\Delta V_K$ and the sixth column the values we find using the BHS interpretation. Cases (i) to (iv) come from the work of Baumgart et al. \cite{baumgart2009quantitative} and demonstrate that the BHS interpretation can predict results that agree with experiment. However, in cases (v) to (x) we find a significant discrepancy between the predictions and the experimental results. Apparently, the BHS interpretation is not valid for these cases.

\begin{table}
\caption{\label{tab:table1}Four KPFM experiments on Si \emph{pn}-junctions from Baumgart et al. \cite{baumgart2009quantitative} and six from other references. From left to right, the columns give the reference, a case label, the reported dopant concentrations, experimental $\Delta V_{K}$ and the predictions for $\Delta V_K$ by the BHS interpretation. Note that the BHS predictions for case (v) to (x) deviate significantly from the experimental values.}
\begin{ruledtabular}
\begin{tabular}{ccccdd}
\multicolumn{1}{c}{Ref.}&\multicolumn{1}{c}{case}&\multicolumn{1}{c}{$N_{A}$[cm$^{-3}$]}&\multicolumn{1}{c}{$N_{D}$[cm$^{-3}$]}&\multicolumn{1}{c}{$\Delta V_{K}^\text{exp}[V]$}&\multicolumn{1}{c}{$\Delta V_{K}^\text{BHS}$[V]}\\
\hline
$[$\onlinecite{baumgart2009quantitative}$]$&(i)&$2 \times 10^{16}$(B)&$2 \times 10^{17}$(P)&0.30&0.309 \\
$[$\onlinecite{baumgart2009quantitative}$]$&(ii)&$2 \times 10^{16}$(B)&$2 \times 10^{20}$(As)&0.20&0.194 \\
$[$\onlinecite{baumgart2009quantitative}$]$&(iii)&$4.7 \times 10^{16}$(B)&$1.4 \times 10^{15}$(P)&0.44&0.411 \\
$[$\onlinecite{baumgart2009quantitative}$]$&(iv)&$1 \times 10^{15}$(B)&$6.5 \times 10^{15}$(P)&0.47&0.469 \\
$[$\onlinecite{saraf2005localSS}$]$&(v)&$1.8 \times 10^{15}$(B)&$2.1 \times 10^{20}$(As)&0.69&0.254 \\
$[$\onlinecite{tsui2008two}$]$&(vi)&$5 \times 10^{14}$(B)&$2 \times 10^{20}$(As)\footnote{$N_{D}$ was extrapolated from Fig. 9 and the given $\Delta V_{K}$ in Ref. \cite{tsui2008two}.}&0.23&0.287 \\
$[$\onlinecite{tsui2008two}$]$&(vii)&$5 \times 10^{14}$(B)&$2 \times 10^{20}$(As)\footnotemark[1]&0.02\footnote{$\Delta V_{K}$ was estimated from Fig. 6(c). in Ref. \cite{tsui2008two}}&0.287 \\
$[$\onlinecite{volotsenko2010secondary}$]$&(viii)&$1 \times 10^{19}$(B)&$3.5 \times 10^{15}$(P)&0.07&0.284 \\
$[$\onlinecite{volotsenko2010secondary}$]$&(ix)&$5 \times 10^{18}$(B)&$3.5 \times 10^{15}$(P)&0.05&0.294 \\
$[$\onlinecite{volotsenko2010secondary}$]$&(x)&$1 \times 10^{18}$(B)&$3.5 \times 10^{15}$(P)&0.03&0.321 \\
\end{tabular}
\end{ruledtabular}
\end{table}

In addition to the erroneous predictions for cases (v) to (x), we identify a more general incorrect behavior of the BHS interpretation. Although the BHS authors mention the importance of surface states, the prediction for a certain dopant concentration does not depend on the amount of surface states or fixed surface charge. For a given \emph{pn}-junction the BHS model therefore predicts a single $\Delta V_{K}$, independent of surface treatment. However, different $\Delta V_{K}$ values for different surface treatments have been reported \cite{sugimura2002potential,tsui2008two}. Cases (vi) and (vii) constitute an example of this. These two cases correspond to samples that have identical \emph{pn}-junctions, as far as dopant concentrations are concerned, but in case (vi) the sample was dipped in HF and not thermally oxidized, while in case (vii) the sample was thermally oxidized and not dipped in HF. The rather different value for $\Delta V_{K}$ measured on these two samples cannot be accounted for by the BHS interpretation.

\begin{figure}[t]
\includegraphics[width = 220pt]{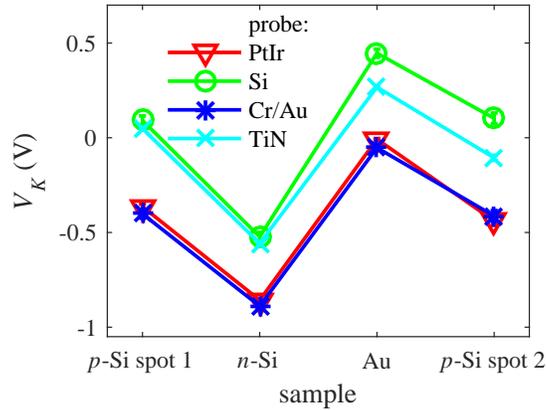}
\caption{\label{KPFM_probe_dep_on_SC} $V_K$ measured with AM-KPFM on different samples with different probes, as indicated in the legend and explained in the text.} 
\end{figure}

Finally, we present evidence that the BHS prediction that KPFM potentials measured on semiconductors should be independent on the probe work function, see Eqs. (\ref{Baumn}) and (\ref{Baump}), is incorrect. This claim is in apparent agreement with the similar results \footnote{We note that in \cite[][Fig. 5]{baumgart2009quantitative} the curves for the \emph{p}- and \emph{n}- type probe practically overlap, but in \cite[][Fig. 6.15(a)]{baumgart2012quantitativethesis} they are shifted by about 150 mV.} they obtained with two highly doped \emph{p}-type and \emph{n}-type probes, which presumably have significantly different work functions. However, we experimentally investigated the probe work function independence for a number of different probes on differently doped Si samples and on Au, and found a clear and reproducible dependence on the probe material, see Fig. \ref{KPFM_probe_dep_on_SC}. Each point in this figure is the mean of four 0.5~$\times$~2~$\mu$m raster scans of 32~lines with 128~pixels. It is generally accepted that in KPFM with a metallic sample and probe, $V_{K}$ is equal to the difference in the work functions of the sample and the probe \cite{nonnenmacher1991kelvin,melitz2011kelvin,sadewasser2012experimental}. Hence, the very similar probe dependence obtained on Au and Si strongly suggests that also on Si KPFM measurements are dependent on the probe work function. This is in accordance with the CPD interpretation and not with the BHS interpretation. We speculate that the nearly identical results obtained with highly doped \emph{p}-type and \emph{n}-type probes by Baumgart et al. were caused by a very high density of surface states at the tip apex, which, through Fermi level pinning, can lead to nearly identical work functions \cite{bardeen1947surface}.

The incorrect predictions of the BHS interpretation described in this section lead us to conclude that it is not universally valid for the interpretation of KPFM data obtained on semiconductors. In the next section, we give support for the correctness of the CPD interpretation and argue that it should be used instead.

\subsection{Testing the CPD interpretation against experiment}
To obtain $\Delta V_K$ from the CPD interpretation, one needs to know the DOSS and the fixed surface charge density. Since these are generally unknown and difficult to measure, we fit the CPD interpretation with the five surface charge models described in section \ref{methods} to all experimental results listed in Table \ref{tab:table1}. Although this disables a stringent test of the CPD interpretation, it turns out that this still gives constraints, due to the fact that the fit parameter of these models (the total surface state density $N_{ss}$ for the DOSS models and fixed surface charge density $\sigma_{sf}$ for the fixed surface charge models) must be of reasonable value (to be discussed below). More importantly, our purpose here is not to subject the CPD interpretation to scrutiny, but rather to demonstrate that – in contrast to the BHS interpretation – the CPD interpretation \emph{is} capable of accommodating the experimental observations discussed in the previous section.

\begin{figure}[t]
\includegraphics[width = 220pt]{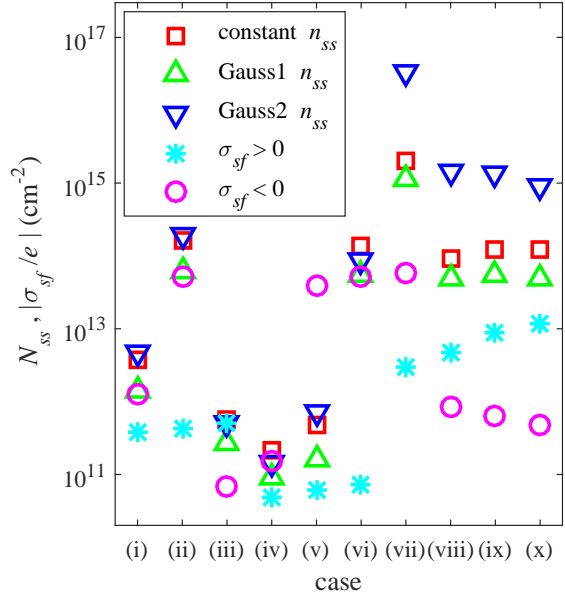}
\caption{\label{fitresults}Fit parameter values obtained by fitting the CPD interpretation to the experimental $\Delta V_K$ listed in Table~\ref{tab:table1} to within 1~mV. The legend indicates the corresponding surface charge model. The models and their labeling are described in section~\ref{CompMethods}.} 
\end{figure}

The values of the fit parameters that reproduce the experimental $\Delta V_K$ values listed in Table \ref{tab:table1} to 1 mV precision are presented in Fig. \ref{fitresults}. We also calculated the sensitivity of the fit parameters by fitting them to the experimental values $\pm$5~mV. It was found that the resulting range falls within the symbols plotted in Fig. \ref{fitresults}. Due to the complicated behavior of the $\widetilde{W}_s$ in the fixed surface charge models, there can be multiple solutions $\sigma_{sf}$ that reproduce a certain value of $\Delta V_K$. However, we checked that within the range $10^{10}$~cm$^{-2}>\left\vert\sigma_{sf}/e\right\vert>10^{14}$~cm$^{-2}$ there is only one solution for each case . 

We consider values of $N_{ss}$ and $\sigma_{sf}/e$ below $~10^{13}$~cm$^{-2}$ to be reasonable, see \cite{deal1967characteristics,flietner1988spectrum}. Higher values are increasingly unlikely with increasing density. Although the actual samples were possibly rough, surface state densities above the density of surface atoms ($\backsim10^{15}$~cm$^{-2}$) are very unlikely for Si surfaces that have been exposed to air. As a result, the main conclusion from Fig. \ref{fitresults} is that all cases can be fit with a reasonable value of the fit parameter by at least one model. This demonstrates that the CPD interpretation \emph{is} capable of accommodating the experimental observations discussed in the previous section. We will now use these fit results to draw some conclusions with respect to the validity of the five surface models for the individual cases.

Case (i) can be fit with all five models with reasonable fit parameter values (i.e. below $~10^{13}$~cm$^{-2}$), while for case (ii) only the fixed positive surface charge model leads to reasonable values; the other models lead to rather high densities. The experimental $\Delta V_K$ values of case (i) and (ii) are obtained from a single KPFM scan on a single sample with multiple \emph{pn}-junctions. Hence, these junctions have undergone similar surface treatments, suggesting that their surface state density or fixed surface charge should be similar. Therefore, since only the fixed positive surface charge model gives nearly identical fit results it is the most likely for both cases. In addition, in case (ii) all other models lead to very high parameter values.

As discussed in the previous section, cases (vi) and (vii) correspond to samples that have identical \emph{pn}-junctions, as far as dopant concentrations are concerned, but which have undergone different surface treatments. In contrast to the BHS interpretation, the CPD interpretation can accommodate different obtained $\Delta V_K$ by a different band bending $V_s$, resulting from a different $N_{ss}$ or $\sigma_{sf}$. Since all models, except the fixed positive surface charge, lead to rather high fit parameter values for these cases, the most likely explanation for the observed difference in $\Delta V_K$ is a higher positive $\sigma_{sf}$ in case (vii) than in case (vi).

Like case (i) and (ii), cases (viii) to (x) correspond to a single Si sample with several \emph{pn}-junctions that went through a single preparation process. For these cases it is therefore again reasonable to assume that the surface state density or fixed surface charge should be similar. Interestingly, this corresponds well with the observation that for each model the fit parameter values for these three cases are very similar. However, the surface state models lead to rather high surface state densities and are therefore less likely. The negative fixed surface charge model is the only that leads to densities that are significantly below $~10^{13}$~cm$^{-2}$, but also the positive fixed surface charge model appears to be reasonable.

In the analysis of the experimental results of cases (viii) to (x), Volotsenko et al. \cite{volotsenko2010secondary} assumed zero band bending, i.e. $V_s=0$, on the highest doped \emph{p}-type region, which is the \emph{p}-side in case (viii). However, in our calculations $V_s$ is larger than 400~mV in this region in all fitted CPD models, except in the negative fixed surface charge model, where it is only $-9$~mV. This suggests that either the assumption was not justified, or there was a fixed negative surface charge. Importantly, both conclusions would significantly influence the results of their further analysis.

Our calculations demonstrate that the CPD interpretation can accommodate all the results discussed in the previous section, even those for which the BHS interpretation gives predictions that do not agree with experiment. We have also shown that, contrary to the BHS interpretation, the CPD interpretation can accommodate different $\Delta V_K$ measured on identical \emph{pn}-junctions that have gone through different surface preparation treatments. And, clearly, the erroneous BHS prediction that $V_{K}$ is independent of the probe work function is absent in the CPD interpretation (see Eq. (\ref{CPDinterp}) and (\ref{CPDWS})). Hence, in agreement with the theoretical arguments given in section \ref{KPprinciples}, we posit that the CPD interpretation is valid for KPFM on semiconductors.

\section{Conclusions}
In this work two different interpretations for KPFM measurements on semiconductors are studied; the CPD interpretation and the BHS interpretation proposed by Baumgart et al. \cite{baumgart2009quantitative,baumgart2011kelvin,baumgart2012quantitativethesis}. By performing model calculations we show that they generally lead to very different results and, thus, that it is important to decide which one should be used.

We show that the BHS interpretation predicts Kelvin potential differences as obtained by KPFM that are not in agreement with experimental observations on Si \emph{pn}-junctions that have been reported in literature. A more general incorrect prediction is that for a specific doping profile, it predicts a KPFM potential difference across a \emph{pn}-junction that is independent of the surface treatment, while some experimental potential differences reported in the literature are very different for different surface treatments. Finally, it predicts that the absolute value of the measured potential is independent of the probe work function, while our own KPFM measurements on Si demonstrate a clear dependence on the probe material. We find that this dependence is very similar to the dependence obtained on Au, which suggests that on semiconductors the absolute value of the measured potential depends on probe work function in the same way as on Au, as predicted by the CPD interpretation. In addition, we show that the CPD interpretation is able to accommodate all the discussed experimental results, including those for which the BHS interpretation gives erroneous predictions.

Based on these findings we posit that the BHS interpretation is not generally suitable for the analysis of KPFM on semiconductors and that the CPD interpretation should be used instead.

\section{Acknowledgements}
We thank Rick Elbersen for help in the sample preparation. This work is part of the research program of the Foundation for Fundamental Research on Matter (FOM), which is part of the Netherlands Organization for Scientific Research (NWO) and was carried out within the research program of BioSolar Cells, co-financed by the Dutch Ministry of Economic Affairs.

\FloatBarrier
%Reference section
\providecommand{\noopsort}[1]{}\providecommand{\singleletter}[1]{#1}%

\end{document}